\documentclass[pdftex,twocolumn,epjc3]{svjour3}          

\usepackage[T1]{fontenc}
\usepackage[english]{babel}
 \addto\captionsenglish{
                        \renewcommand{\figurename} {FIG.}%
                        \renewcommand{\tablename}  {TABLE}%
                        \renewcommand{\appendixname}{APPENDIX}%
                       }
\usepackage[numbers,sort&compress]{natbib} 

\usepackage{graphicx}
 \graphicspath{{graphics/}}
\usepackage{multirow}

\usepackage[caption=false]{subfig}  
\usepackage[export]{adjustbox}

\usepackage{amsmath}
\usepackage{amssymb}
\usepackage[version=3]{mhchem}
\usepackage{mathrsfs}
\usepackage{bm}
\usepackage{todonotes}
\usepackage{upgreek}
\usepackage{soul} 
\usepackage{doi} 
\hypersetup{
            colorlinks,
            linkcolor={red!70!black}  ,
            citecolor={green!70!black},
            urlcolor ={blue!70!black}
           }

\usepackage{comment} 
\usepackage[switch]{lineno} 
\usepackage{xspace}

\newcommand{\LWO}{Li$_2$WO$_4$}
\newcommand{\LMO}{Li$_2$MoO$_4$}
\newcommand{\Thlivetime}{\textcolor{black}{98~}}
\newcommand{\AmBelivetime}{\textcolor{black}{28~}}
\newcommand{\LWOThreshold}{\textcolor{black}{1.5~}}
\newcommand{\LWOBaseline}{\textcolor{black}{$0.50\pm0.01$~}}
\newcommand{\LDBaseline}{\textcolor{black}{$27.2\pm0.1$~}}
\newcommand{\LDFe}{\textcolor{black}{$77 \pm 1$~}}
\newcommand{\LYal}{\textcolor{black}{$0.13 \pm 0.01$~}}
\newcommand{\LYbg}{\textcolor{black}{$0.59 \pm 0.06$~}}
\newcommand{\LYnr}{\textcolor{black}{$0.08 \pm 0.04$~}}

\newcommand{\vet}[1]{\ensuremath{\hskip-1pt\vec{\hskip1pt#1}}}

\begin{document}

 \journalname{Eur. Phys. J. C}
 \title{Characterization of the \LWO\ crystal as a cryogenic scintillating calorimeter}


  \author{D.~L.~Helis~\thanksref{INFNlngs}
         \and A.~Melchiorre~\thanksref{UnivAQ,INFNlngs,corrAuth}
         \and S.~Nagorny~\thanksref{GSSI,INFNlngs,corrAuth}
         \and M.~Noia~\thanksref{GSSI,Zurigo}
         \and L.~Pagnanini~\thanksref{GSSI,INFNlngs,corrAuth}
    	 \and S.~Pirro~\thanksref{INFNlngs}
         \and A.~Puiu~\thanksref{INFNlngs}     
         \and G.~Benato~\thanksref{GSSI, INFNlngs}
         \and P.~Carniti~\thanksref{Mib,INFNMib}
         \and R.~Elleboro~\thanksref{UnivAQ,INFNlngs}
         \and P.~Gambacorta~\thanksref{UnivAQ,INFNlngs}
         \and C.~Gotti~\thanksref{INFNMib}
         \and V.~D.~Grigorieva~\thanksref{Nov}
         \and S.~Nisi~\thanksref{INFNlngs}
         \and E.~Olivieri~\thanksref{IJCLab}
         \and G.~Pessina~\thanksref{INFNMib} 
         \and S.~Piacentini~\thanksref{GSSI, INFNlngs}
         \and M.~Shafiee~\thanksref{Naza}
         \and V.~N.~Shlegel~\thanksref{Nov}
 }
\institute{INFN -- Laboratori Nazionali del Gran Sasso, Assergi, I-67100 L'Aquila,  Italy\label{INFNlngs}
            \and Dipartimento di Scienze Fisiche e Chimiche, Universit\`a degli Studi dell'Aquila, I-67100 L'Aquila, Italy\label{UnivAQ}
            \and Gran Sasso Science Institute, I-67100 L'Aquila, Italy\label{GSSI}
            \and INFN -- Sezione di Milano - Bicocca, I-20126 Milano, Italy \label{INFNMib}
            \and Università degli Studi di Milano -- Bicocca, I-20126 Milano, Italy \label{Mib}
            \and Nikolaev Institute of Inorganic Chemistry, SB RAS, Novosibirsk 630090, Russia \label{Nov}
            \and Université Paris-Saclay, CNRS/IN2P3, IJCLab, 91405 Orsay, France \label{IJCLab} 
            \and Department of Electrical and Computer Engineering, Nazarbayev University, 010000 Astana, Kazakhstan\label{Naza}
            \and Now at: Physik-Institut, University of Z\"urich, 8057 Z\"urich, Switzerland\label{Zurigo}
            \\
         }
           \newcommand{\emailaddress}{}
           \thankstext{corrAuth}{Corresponding Authors: \href{mailto:\emailaddress}{andrea.melchiorre@graduate.univaq.it, serge.nagorny@gssi.it, lorenzo.pagnanini@gssi.it}}
           

 \twocolumn
 \maketitle

\begin{abstract}
A wide range of scintillating bolometers is under investigation for applications in the search for rare events and processes beyond the Standard Model. In this work, we report the first measurement of a natural, non-molybdenum-doped, lithium tungstate (\LWO) crystal operated underground as a scintillating cryogenic calorimeter. The detector achieved a baseline energy resolution of 0.5~keV RMS with a low-energy threshold of about \LWOThreshold keV. The simultaneous readout of heat and light enabled particle identification, revealing a clear separation between $\beta/\gamma$, $\alpha$, and nuclear recoil populations above 300 keV, with a light-yield–based particle discrimination better than $6\sigma$. These results, fully comparable with those achieved with other compounds in the field, demonstrate that \LWO\ is a promising candidate for rare-event searches. In particular, the combination of excellent radio-purity (with U/Th levels below 0.5~mBq/kg) and sensitivity to neutron interactions via the $^6$Li(n,$\alpha$)$^3$H reaction makes this material an attractive option for next-generation experiments on dark matter, coherent elastic neutrino-nucleus scattering, and spin-dependent interactions.
 \end{abstract}
 
 
\section{Introduction}
Coherent Elastic Neutrino-Nucleus Scattering (CE$\nu$NS) process represents a unique probe at the intersection of particle, nuclear, and astroparticle physics. Predicted within the Standard Model (SM)~\cite{Freedman:1973yd,Cadeddu_2023} 50 years ago, CE$\nu$NS involves the scattering of low-energy neutrinos off entire nuclei via neutral-current interactions, resulting in a detectable nuclear recoil with energies ranging from a few eV up to several keV, depending on the target mass. Despite its early theoretical prediction, the process was experimentally observed only in 2017 by the COHERENT collaboration~\cite{COHERENT:2017ipa}, due to the need for ultra-low threshold detectors and the lack of suitable technology until recent advancements~\cite{Schumann:2019eaa}.
The relatively large interaction cross-section of CE$\nu$NS, along with its equal sensitivity to all neutrino flavors via the $Z^0$ boson exchange, makes it an ideal detecting mechanism for neutrinos from various sources, such as the Sun, supernovae, nuclear reactors, and radioactive decays. Its sensitivity to subtle deviations from SM predictions enables the exploration of exotic neutrino properties, non-standard interactions, sterile neutrinos, neutrino magnetic moments, and potential light mediators like new gauge bosons~\cite{Miranda:2019skf,Atzori_Corona_2025,Corona_2025,Corona_2024}. Furthermore, CE$\nu$NS plays a critical role in characterizing neutrino-induced backgrounds for next-generation dark matter experiments, which share the need for low-threshold and high-precision detection.

The CE$\nu$NS differential cross-section of a neutrino with a spin-zero nucleus $\mathcal{N}$ according to the SM principles~\cite{Freedman:1973yd,Cadeddu_2023} is 
\begin{equation}
    \label{eq:xsec}
\begin{split}
\frac{d\sigma}{d E_R} = & \frac{G^2_F m_N}{8 \pi (\hslash c )^4} \cdot \left(2- \frac{E_R m_N}{E_\nu^2}\right)\ \times \\
&\left[(4\sin^2 \theta_W -1 )\cdot Z\cdot F_{Z}^{\mathcal{N}}(\vet{q}) + N\cdot F_{N}^{\mathcal{N}}(\vet{q})\right]^2
\end{split}
\end{equation}
where $G_F$ is the Fermi coupling constant, $\theta_W$ is the Weinberg angle, $Z$ and $N$ are the atomic and neutron numbers of the target nucleus, $m_N$ is its mass, and $E_\nu$ is the incoming neutrino energy.
The functions $F_{N}^{\mathcal{N}}(\vet{q})$ and $F_{Z}^{\mathcal{N}}(\vet{q})$ are, respectively, the neutron and proton form factors of the nucleus $\mathcal{N}$, which characterize the coherency of the interaction and depend on the neutron and proton distributions in the nucleus. 

A strong correlation exists between the incoming neutrino energy ($E_\nu$) and the recoil energy, as shown by the average recoil~\cite{Drukier:1984vhf}:
\begin{equation}
\label{eq:eravg}
\langle E_R \rangle = \frac{2E_\nu^2}{3m_N},
\end{equation}
indicating that higher-energy neutrinos significantly enhance the recoil signal. 

\begin{figure}
    \centering
    \includegraphics[width=\linewidth]{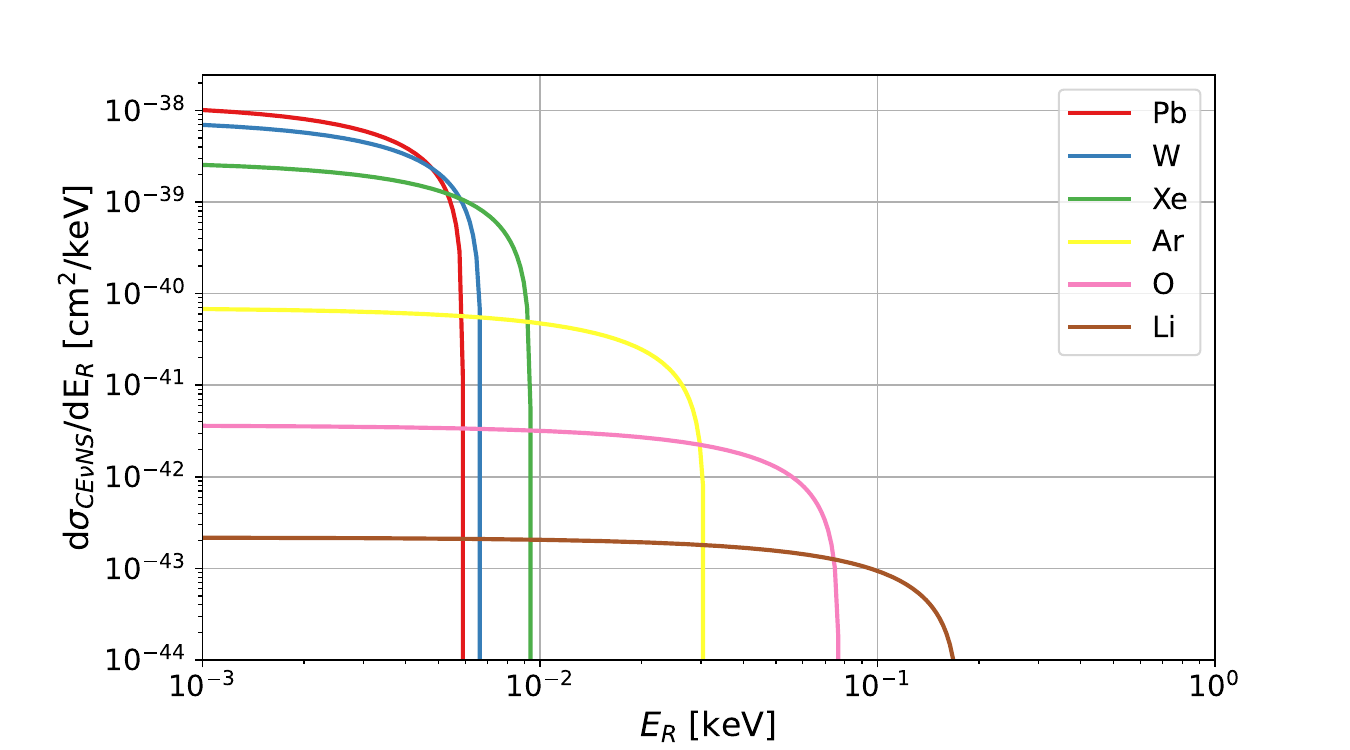}
    \includegraphics[width=\linewidth ]{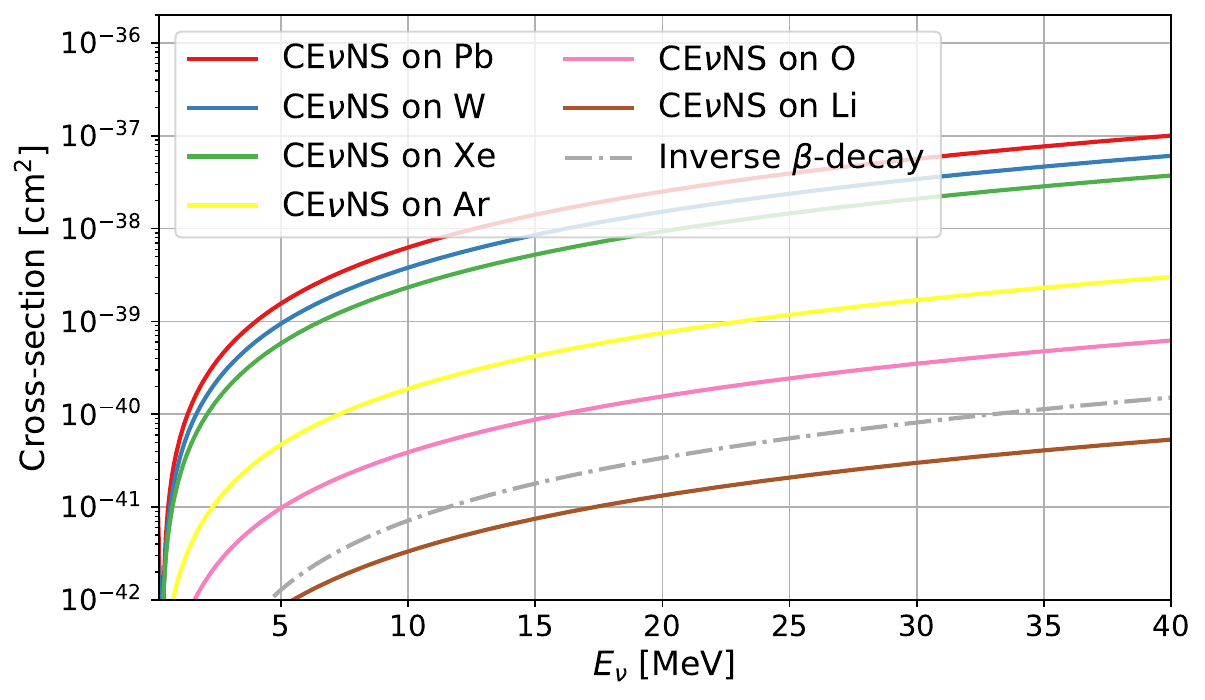}
    \caption{{\it Top.} Coherent elastic neutrino-nucleus scattering (CE$\nu$NS) cross-sections as a function of incoming neutrino energy for various target nuclei. Due to their larger neutron number, heavier elements such as tungsten (W) and lead (Pb) exhibit significantly higher cross sections than lighter nuclei.
    {\it Bottom.} CE$\nu$NS differential cross sections as a function of the nuclear recoil energy. Following the work in Ref.~\cite{Bellenghi:2019vtc}, a monochromatic neutrino of 747~keV emitted in the $^{51}$Cr decay is assumed, while considering a reactor neutrino with an energy of 3-4~MeV, the spectra shift toward higher energies by a factor of $\sim$20 (see Eq.~\ref{eq:eravg}). Lighter elements, such as lithium (Li) and oxygen (O), have a higher energy recoil, thereby relaxing the energy threshold requirements. Figures adapted from Refs.~\cite{RES-NOVA:2021gqp,Alloni:2025bda}.}
    \label{fig:cross_section}
\end{figure}

As shown in Fig.~\ref{fig:cross_section}, CE$\nu$NS cross-sections can be up to $10^3$–$10^4$ times larger than other conventional neutrino detection channels. Scaling with $N^2$ under coherent conditions, heavy nuclei are ideal candidates due to their high neutron number and nuclear stability, enabling both high interaction rates and low backgrounds. Additionally, the cross-section is proportional to the square of the neutrino energy, $E_\nu^2$, further favoring higher energy sources.
For heavy elements (Xe, W, Pb), the cross-section is higher, but the required threshold is lower. Conversely, for light elements (Li, O), the cross-section is lower, but the recoils reach higher energies. Indeed, having a recoil at higher energy can ease the experimental constraint on having an extremely low threshold detector.

The renewed interest in CE$\nu$NS as a SM probe and neutrino detection channel~\cite{Pattavina:2020cqc, Pattavina_2021,Solar_Xe, Solar_Panda} pushes the scientific community to develop detection techniques to probe low-energy nuclear recoils produced by artificial neutrino sources. 
The COHERENT experiment achieved the first detection of coherent elastic neutrino–nucleus scattering at a 6.7 $\sigma$ significance using a 14.6 kg CsI[Na] detector at the Spallation Neutron Source (Oak Ridge National Laboratory, USA), confirming Standard Model predictions and setting new limits on possible nonstandard neutrino interactions~\cite {COHERENT:2017ipa}.
Recently the CONUS+ experiment, based on high-purity germanium detectors with extremely low-energy threshold (160 eV), has achieved the first observation of coherent elastic neutrino–nucleus scattering from reactor neutrinos with a statistical significance of 3.7$\sigma$~\cite{Ackermann:2025obx}.
NUCLEUS~\cite{NUCLEUS:2019igx} employs cryogenic bolometers made of CaWO$_4$ and Al$_2$O$_3$ to detect CE$\nu$NS from continuous flux of reactor neutrinos with thresholds of a few tens of eV. Finally, RICOCHET~\cite{Ricochet:2023nvt} uses cryogenic detectors with Ge and Zn targets to measure CE$\nu$NS from reactor neutrinos, aiming for high precision and sensitivity to new physics. Since these detectors typically operate near nuclear reactors at shallow depths, they must meet a combination of stringent requirements:
\begin{itemize}
    \item compact detector size, to handle the high event rates at surface locations;
    \item high radio-purity, to suppress background from radioactive decays and minimize false nuclear recoil signals;
    \item energy threshold in the eV range, enabling sensitivity to the recoil energies expected from CE$\nu$NS;
    \item particle identification capability, to discriminate nuclear recoils from electron recoils and other backgrounds;
    \item in-situ neutron monitoring capability, to provide valuable information for background characterization and its rejection.
\end{itemize}

From a detection technology perspective, cryogenic calorimeters represent a promising solution to meet all these requirements. 
Operating at milliKelvin temperatures, these detectors can accurately measure extremely small energy deposits~\cite{Pirro:2017ecr,Beretta:2021apz,CUORE:2021ctv}. 
In particular, they achieve energy thresholds at the eV scale, offer excellent energy resolution (as low as 0.1$\%$), and provide efficient particle discrimination between nuclear and electron recoils, particularly through light readout in scintillating crystals. 
All these features make them well-suited for detecting the faint signals characteristic of CE$\nu$NS. Moreover, the versatility in the choice of absorber material in cryogenic calorimeters allows one to choose the most suitable one for the specific physics case. 
A lithium tungstate (\LWO) crystal emerges as a promising candidate among potential target materials for CE$\nu$NS detection. Its heavy tungsten content enhances the interaction rate of CE$\nu$NS (see Fig.~\ref{fig:cross_section}), while its lithium component —with approximately 8\% natural abundance of $^6$Li— enables efficient neutron background suppression via neutron capture reactions. 
Additionally, \LWO\ exhibits scintillation properties~\cite{Cova:2025uka} that enable simultaneous phonon and light detection, which is crucial for distinguishing between $\alpha$ particles, nuclear and electronic recoils at low energies. The interest in this material has grown significantly in recent years, also driven by the establishment of the lithium molybdate (\LMO) as a target material for CUPID~\cite{CUPID:2025avs} and AMORE~\cite{AMoRE:2024loj} in the search for neutrinoless double beta decay of $^{100}$Mo. For instance, in Ref.~\cite{Cova:2025uka}, the scintillation properties of natural \LWO\ and \LMO\ crystals were compared in a wide temperature range.

The first cryogenic characterization of an 8-gram \LWO\ crystal doped with $5\%$ molybdenum was performed in the framework of the BASKET project~\cite{Aliane:2020dbc}. Here the authors demonstrated an excellent baseline noise of 1.1~keV FWHM, an energy resolution of 3.7~keV FWHM at 609~keV, and a clear particle discrimination capability at a level of $5 \sigma$ between $\alpha$ and $\beta/\gamma$ events when assisted by the Neganov–Trofimov–Luke amplification for light detection. These results, together with a low level of internal radioactive contamination, confirm the promising potential of \LWO(Mo) for rare event searches and CE$\nu$NS experiments in aboveground conditions. A further measurement of the same crystal sample was then carried out using a Metallic Magnetic Calorimeter (MMC) as a thermal sensor~\cite{Mauri:2022zom}, achieving an excellent baseline energy resolution of 194.1~eV FWHM ($\sigma=83$~eV). 

This work presents the first measurement of a large-mass non-molybdenum-doped \LWO\ crystal operated underground as a cryogenic scintillating calorimeter to enable light-assisted particle discrimination. The calorimetric detector performance and scintillation light yield of the crystal were investigated to evaluate its potential for rare event detection. These results lay the groundwork for the future use of \LWO-based detectors in next-generation dark matter and neutrino physics experiments.

\begin{figure}
    \centering
    \includegraphics[width=\linewidth]{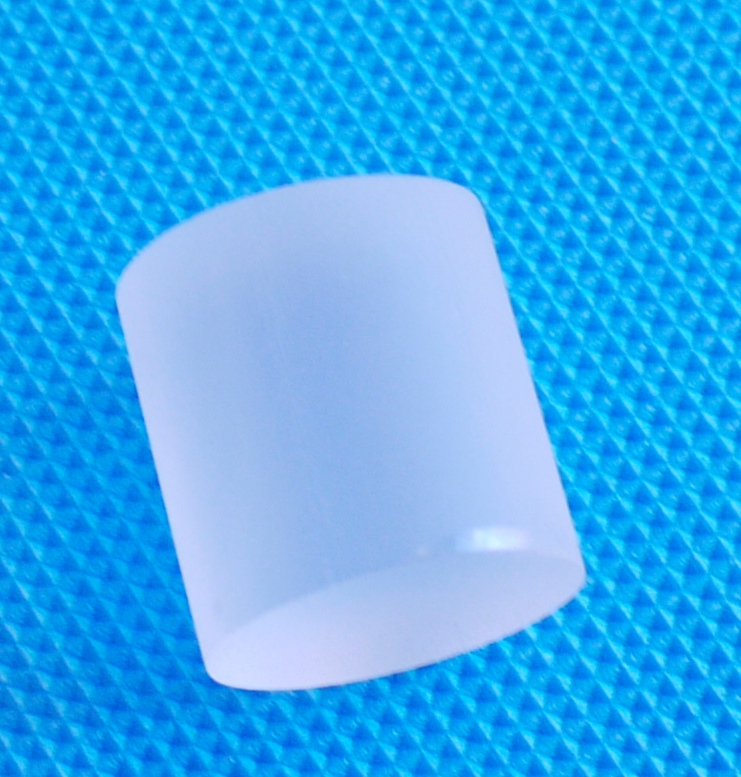}
    \caption{A 24~g \LWO\ crystal with a size of $\varnothing$24.3$\times$25.3~mm before its assembly. }
    \label{fig:set}
\end{figure}

\section{Crystal production}
The \LWO\ single crystal used in this study was grown by a Low-Thermal-Gradient modification (LTG) of the Czochralski technique developed at the Institute of Inorganic Chemistry SB RAS for production of large-volume high-quality oxide crystals~\cite{pavlyuk1992low,borovlev2001progress,arnaboldi2010cdwo4,shlegel2017recent}, and that later was adopted to produce unique crystals for fundamental studies, including isotopically enriched crystals $^{106}$CdWO$_{4}$, $^{116}$CdWO$_{4}$, Zn$^{100}$MoO$_{4}$ and Li$_{2}$$^{100}$MoO$_{4}$~\cite{barabash2011low,belli2010development,grigorieva2020li2100deplmoo4,Grigorieva:2018qaz}.

The initial charge (\LWO\ compound) was obtained through a solid-state synthesis from the commercial lithium carbonate (Li$_{2}$CO$_{3}$) and deeply purified tungsten oxide powder (WO$_{3}$) \cite{Grigorieva:2018qaz} following the reaction:
\begin{equation}
\text{Li}_{2}\text{CO}_{3} + \text{WO}_{3} \rightarrow \text{Li}_2\text{WO}_4 +\text{CO}_{2} \uparrow  
\end{equation}
The thoroughly mixed precursors were placed in a platinum crucible, heated to 450$^{\circ}$C at a rate of 25$^{\circ}$C/h under an air atmosphere, annealed for 10~hours, and then cooled to room temperature at a rate of 100$^{\circ}$C/h. 
For the subsequent crystal growth, the crucible with the charge was heated to 770$^{\circ}$C at a rate of 100$^{\circ}$C/h, held for 4~hours to homogenize the melt, and then cooled slightly to 735$^{\circ}$C for the seeding process. The optimal seed orientation was defined as [001]. The growth parameters (pulling rate of 1~mm/h and rotation rate of 5~rpm) allowed the realization of a standard growth mechanism with a rounded, slightly convex crystallization front, and to obtain the \LWO\ single crystalline ingot without visible defects or inclusions with dimensions 70~mm in length and 30~mm in diameter. Then, a crystal sample with dimensions $\varnothing$24.3$\times$25.3~mm was procured, as shown in Fig.~\ref{fig:set}. One face of the cylindrical sample was optically polished, while the other face and the lateral surface were matted. 


\section{Experimental setup}
The measurements with the dual-readout \LWO\ scintillating calorimeter were performed in the IETI facility, a pulse tube-assisted dilution refrigerator located in Hall~C of the underground Gran Sasso National Laboratory of INFN (Assergi, Italy). The \LWO\ crystal was equipped with a germanium neutron transmutation doped thermistor (Ge-NTD)~\cite{Haller1984}, with dimensions of $3\times3\times1$ mm$^3$, glued on the upper surface with six spots of bi-component epoxy resin (Araldite Rapid). 
A copper holder secures the target crystal, held in place by polytetrafluoroethylene (PTFE) supports.
To detect the scintillation light, a germanium light detector (Ge-LD) is installed facing the upper polished surface of the \LWO\ crystal. 
The Ge-LD consists of a double-sided polished wafer ($\varnothing 44.5$ mm, thickness 170~$\mu$m) with one side coated with an anti-reflective coating (facing the \LWO) made of SiO to increase the light absorption efficiency. 
On the opposite surface of the Ge/LD, a $3\times1\times1$ mm$^3$ Ge-NTD was glued in the same manner. An extensive description of the Ge-LD development and performance is reported in Refs.~\cite{Beeman:2013zva,Barucci:2019ghi}.
The surfaces of the \LWO\ crystal, except the one facing the Ge-LD, were wrapped with a 3M VIKUITI reflective foil.
The Ge-NTDs are connected to the bias and readout system via 25~$\mu$m gold wires, which are grafted onto twisted pairs of constantan wires and then attached to a thermally anchored board located on the mixing chamber plate of the cryostat. Through the wiring resident in the cryostat, the signal reaches room-temperature electronics, which include low-noise front-end boards and DC coupling, along with high-resolution digitizers~\cite{Arnaboldi_2018,Carniti:2022coa}.

During data collection, the Ge-LD continuously faces a \textsuperscript{55}Fe X-ray source directly installed within the detector holder for energy calibration.
To set the energy scale of the \LWO\ detector, an external $\gamma$ source of \textsuperscript{232}Th was employed, positioned between the 300~K shield of the cryostat and the external lead and copper shielding.
In two specific runs, to study the response of the \LWO\ detector to $\alpha$ particles and nuclear recoils, the \textsuperscript{232}Th source was replaced by an Americium-Beryllium (AmBe) neutron source. 
The continuous data stream was registered with a sampling rate of 10~kHz for the \textsuperscript{232}Th and 5~kHz for the AmBe data, applying a low-pass Bessel filter with a cutoff frequency of 500~Hz to reduce higher-frequency noise~\cite{Carniti:2022coa}.
Data were collected for approximately \Thlivetime~hours in this configuration, followed by an additional \AmBelivetime~hours (14 with a water moderator and 14 without it) measurement using an AmBe neutron source.
Data were stored as binary files for offline processing and analysis.

\section{Data Analysis}
\begin{figure}
    \centering
    \includegraphics[width=\linewidth]{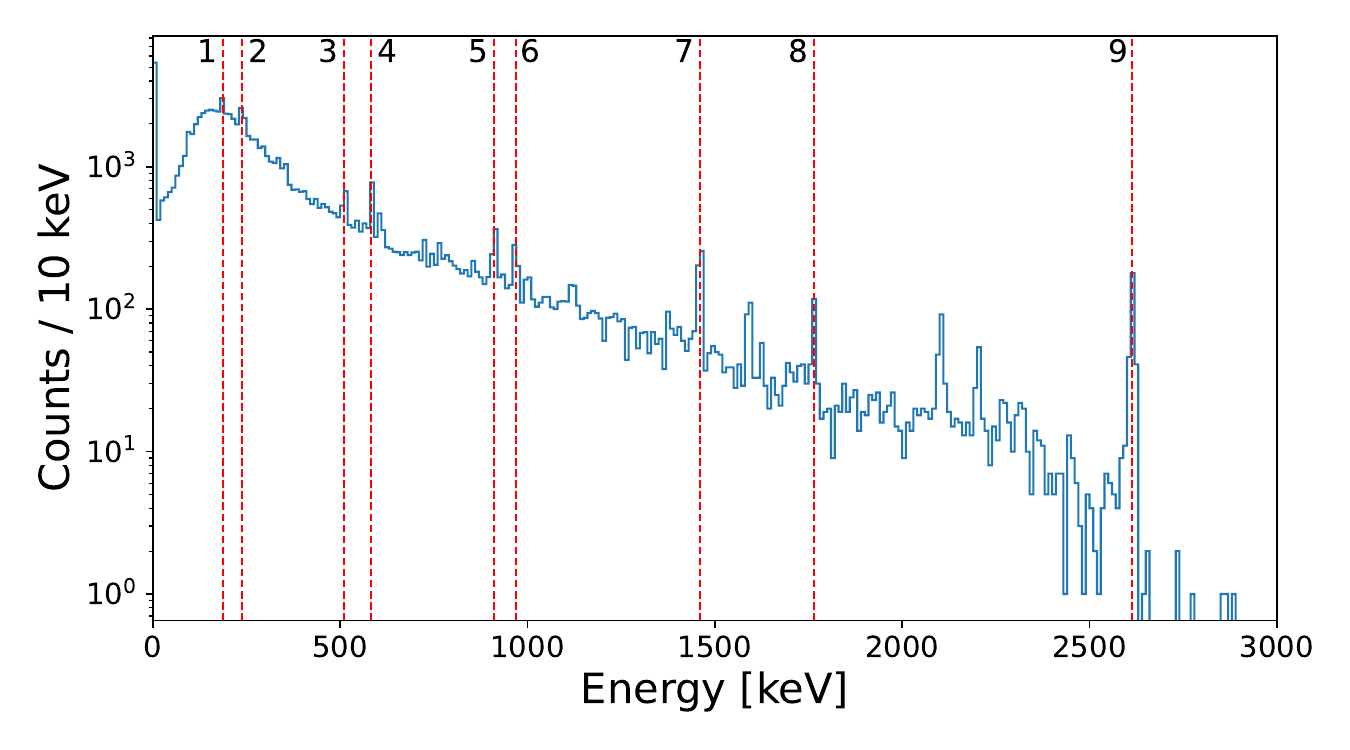}
    \includegraphics[width=\linewidth]{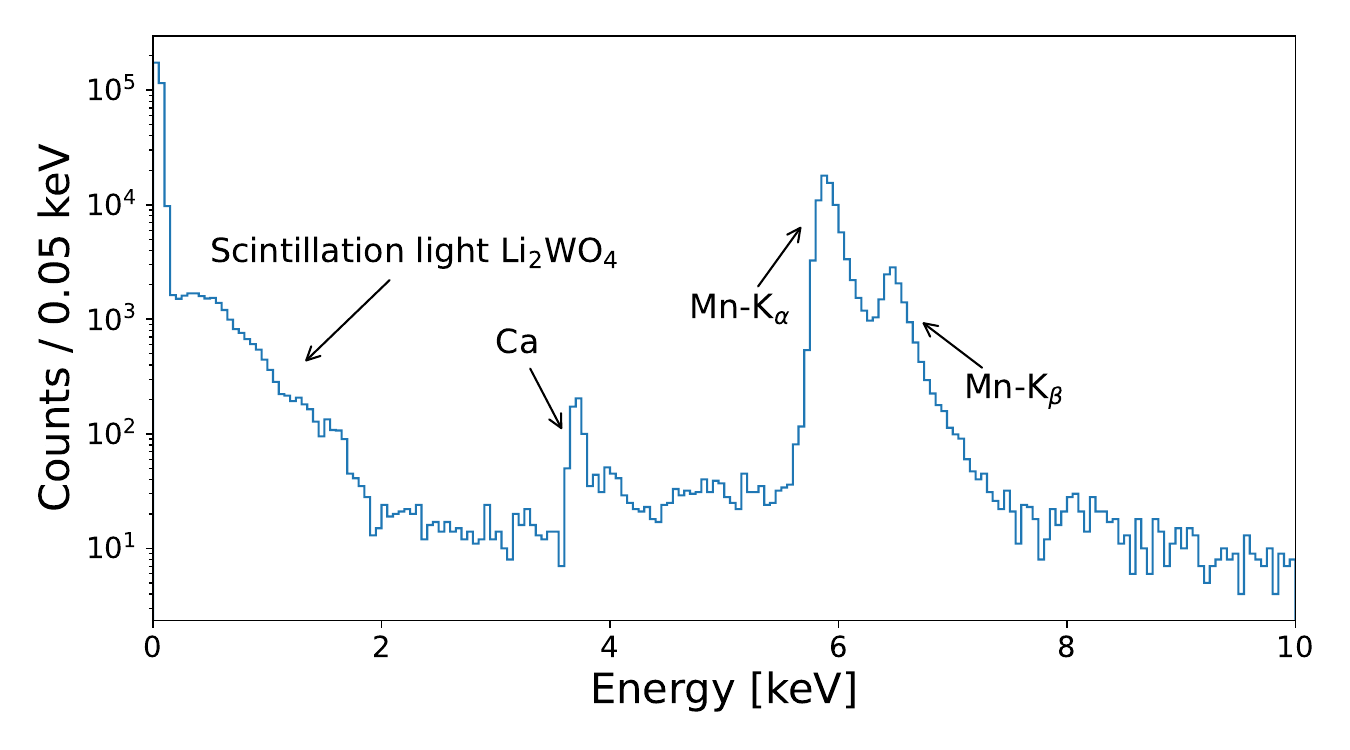}
    \caption{Energy spectra acquired over 98 hours with the \LWO\ crystal (top) and the Ge-LD (bottom) using an external \textsuperscript{232}Th calibration source. In the \LWO\ spectrum, all peaks exploited for the energy calibration are highlighted and labeled following the list in Tab.~\ref{tab:energy_resolution}. Due to its small thickness (170~$\mu$m), the light detector cannot fully absorb the \textsuperscript{232}Th $\gamma$-rays, resulting in a featureless continuum spectrum. Superimposed on this latter are the scintillation light signals from the \LWO\ crystal (0--2~keV), as well as X-rays from the $^{55}$Fe calibration source (5.9 and 6.4~keV), and characteristic calcium X-rays (3.7 and 4.0~keV).}
    \label{fig:spec}
\end{figure}

The collected data were analyzed using a custom software framework designed to trigger and process signals from cryogenic calorimeters.
An event was flagged as a signal if its amplitude exceeded a threshold of 3$\sigma$ above the baseline noise. 
When a trigger fires on the \LWO, the corresponding light signal is automatically triggered on the Ge-LD even if below threshold. In addition, noise events are randomly selected every 5~seconds for characterization purposes. 
A multi-step analysis procedure was employed to reconstruct the amplitude of triggered events.
First, the baseline level, slope of the pre-trigger, root mean square (RMS) noise, rise time, decay time, and unfiltered amplitude within a predefined analysis window were computed.
Then, the signal and noise waveforms were selected to construct the optimal filter transfer function~\cite{Gatti:1986cw}, considering only events with a single trigger per acquisition window and a stable pre-trigger baseline.  
For the \LWO\ detector, the signal template is derived from events at 2.6~MeV to ensure a high signal-to-noise ratio. 
After applying the optimum filter, thermal gain stabilization was performed to correct for possible amplitude drifts, using events from $^{208}$Tl at 2.6~MeV.
The same analysis procedures were applied to the Ge-LD, using events from the \textsuperscript{55}Fe peak at 5.9~keV to perform both stabilization and calibration. 
The energy scale of the heat channel was calibrated using nine prominent spectral peaks from \textsuperscript{232}Th.
Figure~\ref{fig:spec} shows the resulting  calibration spectrum of \textsuperscript{232}Th collected over 98~hours with the \LWO\ crystal (top panel) and the Ge-LD energy spectrum calibrated with \textsuperscript{55}Fe (bottom panel). These spectra were obtained by removing spurious events based on the following criteria:
\begin{itemize}
    \item events with multiple triggers within the analysis window were rejected;
    \item only waveforms with a baseline slope within $\pm$2$\sigma$ of the distribution were retained;
    \item rise and decay times were required to fall within $\pm$3$\sigma$ of their mean values.
\end{itemize}

More details on the data processing are reported in Refs.~\cite{Melchiorre:2025wue,Helis:2024vhr}.

\section{Results}
\subsection{Energy Threshold and Energy Resolution}
Following the above-described analysis procedure, a \LWOBaseline keV RMS baseline noise was achieved, referring to  
\LWOThreshold keV trigger threshold for the \LWO\ detector, defined as 3$\sigma$ of the baseline noise.
The energy resolution of the \LWO\ detector was evaluated using the calibration $\gamma$-ray peaks, as summarized in Table~\ref{tab:energy_resolution}.
For the LD, the baseline resolution of \LDBaseline eV RMS was measured, along with the energy resolution of $\sigma=$~\LDFe eV at 5.9~keV. A global summary of the main performance is reported in Table~\ref{tab:perf_summary}.

Although still far from the requirements for CE$\nu$NS detection, these results demonstrate 
the feasibility of using the molybdenum-free \LWO\ crystal as a cryogenic calorimeter for radiation detection. 

\begin{table}
   \centering
   \caption{Energy resolution of the \LWO\ detector at various energies corresponding to the calibration peaks. For completeness, the resolutions of the baseline and of the 59.5 keV line due to AmBe source are also reported.}
    \begin{tabular}{lccc}
    \hline
    \# & Isotope & Energy & Resolution ($\sigma$)\\ 
    &  & (keV) & (keV)  \\
    \hline
    - & - & Baseline (Noise) & \LWOBaseline  \\
    - & $^{241}$Am & 59.5  & 0.9 $\pm$ 0.2 \\
    1 & $^{226}$Ra & 186.2  & 1.3 $\pm$ 0.1 \\
    2 & $^{212}$Pb & 238.6  & 1.7 $\pm$ 0.1  \\
    3 & $^{208}$Tl + e$^{+}$e$^{-}$ &  510.7 +511  & 2.3 $\pm$ 0.2 \\
    4 & $^{208}$Tl & 583.2  & 1.9 $\pm$ 0.2  \\
    5 & $^{228}$Ac & 911.2  & 2.5 $\pm$ 0.2  \\
    6 & $^{228}$Ac & 964.8/969.0  & 2.6 $\pm$ 0.2 \\ 
    7 & $^{40}$K & 1460.8 & 3.2 $\pm$ 0.2  \\
    8 & $^{214}$Bi & 1764.5 & 3.7 $\pm$ 0.4 \\
    9 & $^{208}$Tl & 2614.5 & 4.3 $\pm$ 0.4  \\
    \hline
   \end{tabular}
   \label{tab:energy_resolution}
\end{table}

\begin{table}[ht!]
   \centering
    \caption{Summary of the detector features and performance. The quoted uncertainties are calculated from the Gaussian fit of the parameter distribution as the uncertainty on the mean value. Only the light yield (LY) uncertainty is quoted as the $\sigma$ of the distribution.}
    \begin{tabular}{l c l}
    \hline
    Parameter & Value & Unit  \\
    \hline
    \textbf{LWO Detector:} &  & \\
    Mass & 24 & [g]\\
    Diameter &  24.3 & [mm]\\
    Height & 25.3 & [mm]\\
    Density & 6.45 & [g/cm$^3$]\\
    Rise Time (10-90\%) & 11 $\pm$ 3 & [ms]\\
    Decay Time (90-30\%) & 52 $\pm$ 2 & [ms]\\
    Signal Amplitude & 53 &[$\mu$V/MeV]\\
    Energy Resolution ($\sigma$) & \LWOBaseline & [keV]\\
    \hline
    \textbf{LD Detector:} &  & \\
    Mass & 0.5 & [g]\\
    Diameter & 2.54  & [cm]\\
    Thickness & 170 & [$\mu$m]\\
    Rise Time (10-90\%) & 1.2 $\pm$ 0.1 & [ms]\\
    Decay Time (90-30\%) & 1.0 $\pm$ 0.1 & [ms]\\
    Signal Amplitude &8228&[$\mu$V/MeV]\\
    Baseline Resolution (RMS) & 27.2 $\pm$ 0.1  & [eV]\\
    Energy Resolution at 5.9 keV & 77 $\pm$ 1 & [eV]\\
    \hline
    LY$_{\beta/\gamma}$ at (0.3 - 3) MeV & \LYbg & [keV/MeV]\\
    LY$_{\alpha}$ at (5 - 6) MeV &  \LYal & [keV/MeV]\\
    LY$_{NR}$ at (0.3 - 2.5) MeV & \LYnr & [keV/MeV]\\
   \hline
   \end{tabular}
   \label{tab:perf_summary}
\end{table}

\subsection{Particle identification in \LWO}
To investigate the particle discrimination capability of the \LWO\ detector at low-energy, a neutron calibration was performed using an AmBe source. The latter produces fast neutrons through the nuclear reaction
\begin{equation}
\alpha + \ce{^9Be} \rightarrow \ce{^{12}C}^{*} + \ce{n}   
\end{equation}
with neutron energies ranging approximately from 1 to 10~MeV. 
Since fast neutrons interact primarily via elastic scattering, they induce nuclear recoils of lithium, tungsten, and oxygen within the \LWO\ crystal. In addition to neutrons, the source emits high-energy $\gamma$-rays, predominantly at 4.4~MeV, resulting from the de-excitation of $^{12}$C nuclei. The AmBe source was placed outside the cryostat in two different configurations: with and without a water moderator. Data was taken over 14 hours in each configuration. 

\begin{figure}
    \centering
    \includegraphics[width=\linewidth]{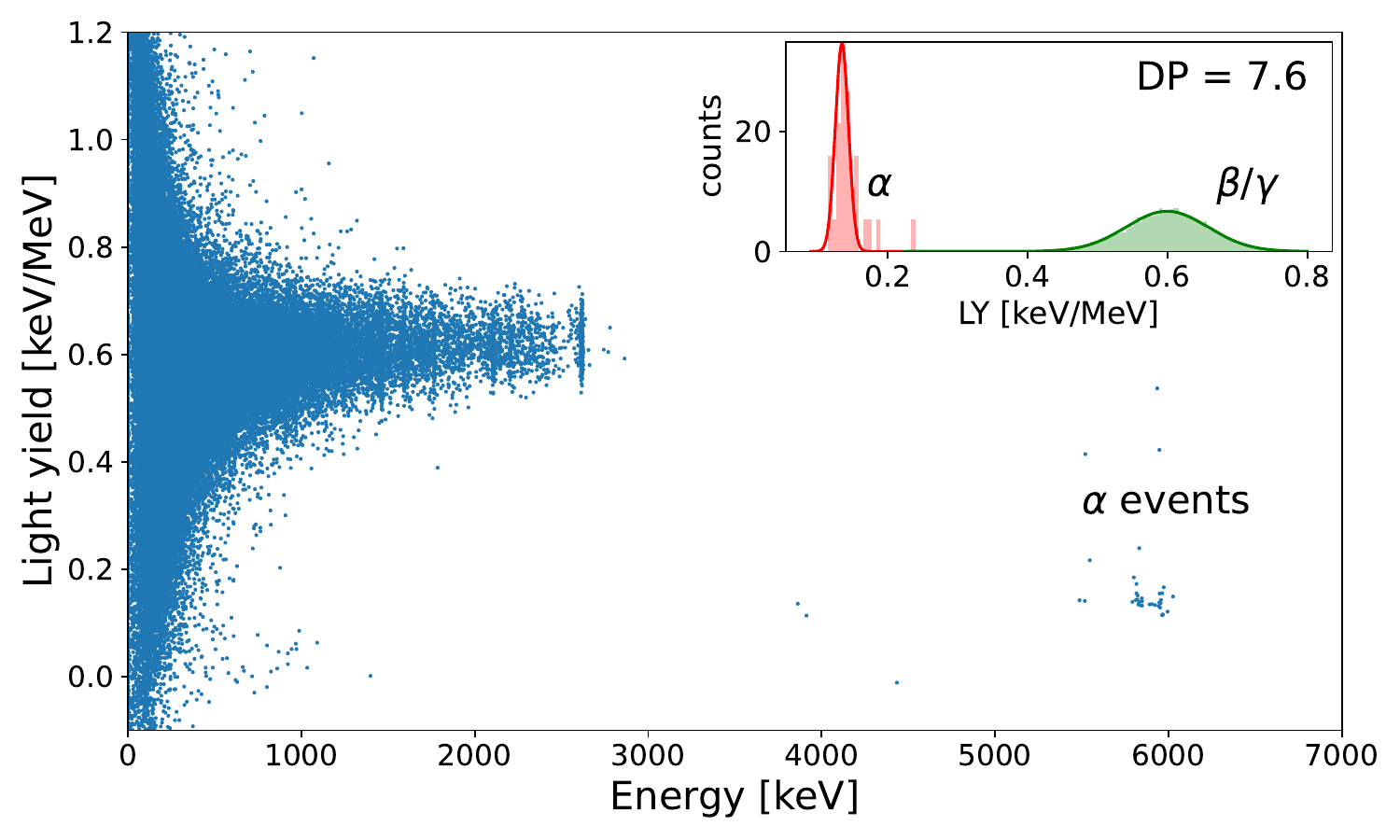}
    \includegraphics[width=\linewidth]{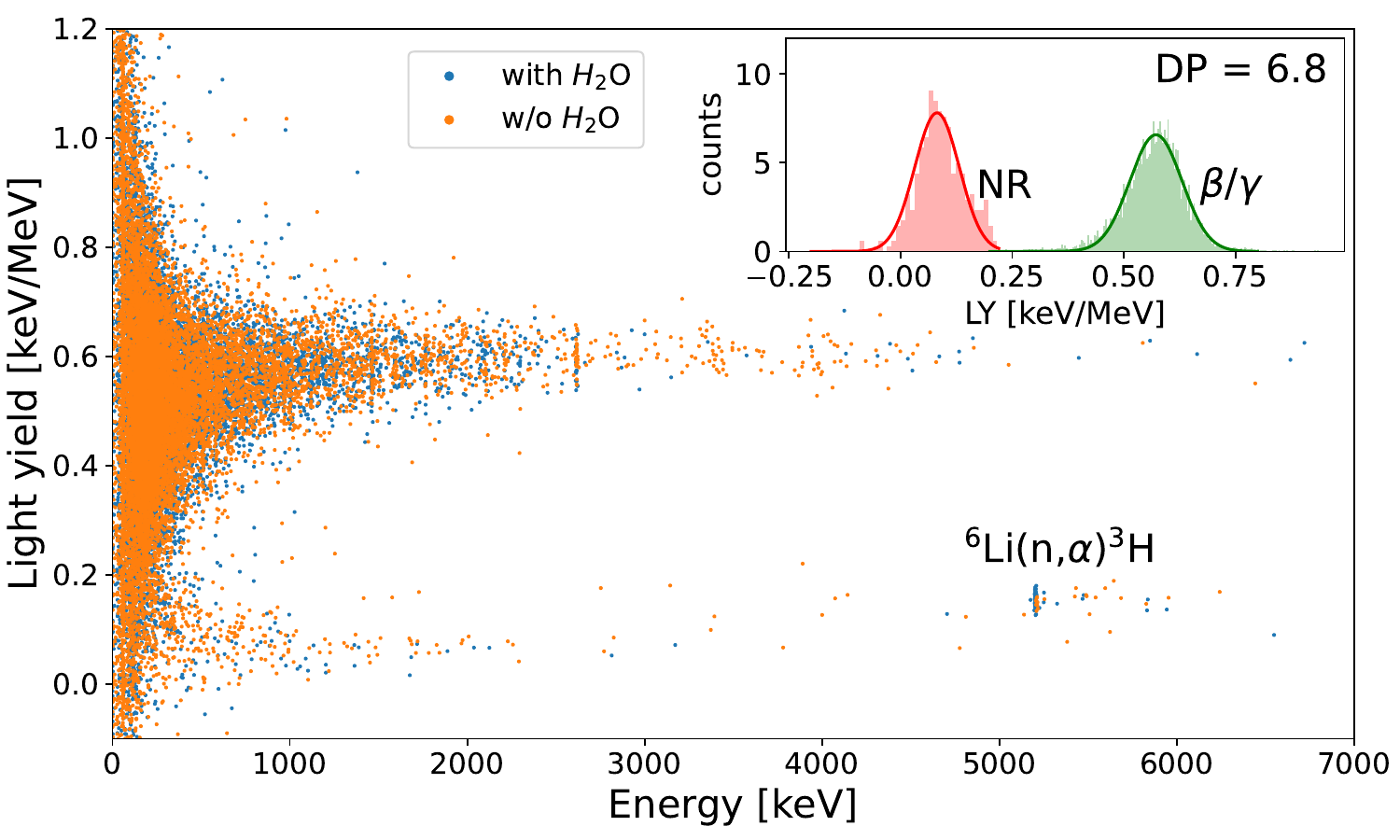}
    \caption{Light yield as a function of the energy for the data acquired with
    \textsuperscript{232}Th source (top panel) and AmBe neutron source (bottom panel).
    In the latter case, both the data acquired with (blue) and without (orange) the water tank used as neutron moderator are reported. Four populations of events are clearly distinguishable in the two datasets: the $\beta/\gamma$s forming a continuous band at LY $\sim 0.6$~keV/MeV, the $\alpha$ particles due to crystal contamination (top) and neutron capture on $^6$Li (bottom) at LY $\sim 0.1$~keV/MeV, as well as the neutron-induced nuclear recoils (NR) at LY $\sim 0.08$~keV/MeV (bottom). The insets report the LY of different particles, exploited to evaluate the discrimination power. See the main text for more details.}
    \label{fig:scatterAmbe}
\end{figure}

Fig.~\ref{fig:scatterAmbe} presents the resulting scatter plot of the scintillation light yield versus energy. It features a prominent $\alpha$ peak at 5190~keV, corresponding to neutron capture events on $^{6}$Li. Notably, the nominal energy of this process is 4784~keV, and an 8\% shift due to thermal quenching is observed, consistent with previous measurements in similar crystals~\cite{CUPID:2022opf,CUPID:2020itw}. The energy resolution of this peak is evaluated to be $\sigma = 3.5 \pm 0.5$~keV.

The scintillation light yield (LY), defined as the ratio of the light signal to the heat signal produced at a certain energy deposition, was used as a key parameter for particle identification. Expressed in keV/MeV, the detected LY is influenced by several experimental factors, including scintillation efficiency of the crystal at low temperatures, its optical properties (i.e. optical transmittance), crystal geometry and its surface treatment, light collection efficiency of the setup, and the Ge-LD light detection efficiency. For these reasons, what is quoted in the following is not the absolute LY of the \LWO\, but the LY measured in this specific detector configuration.

As shown in Fig.~\ref{fig:scatterAmbe}, four distinct event populations are observed: 
\begin{itemize}
    \item $\beta/\gamma$ band, with a LY of \LYbg keV/MeV evaluated in the range 0.3 -- 3~MeV on \textsuperscript{232}Th data;
    \item $\alpha$ particles, with a LY of \LYal keV/MeV evaluated at 2 -- 6~MeV on \textsuperscript{232}Th data;
    \item $\alpha$ particles $+$ $^3$H, due to neutron capture on $^6$Li, with a LY of \LYal keV/MeV evaluated at 5 -- 6~MeV on AmBe data;
    \item nuclear recoils, with a LY of \LYnr keV/MeV evaluated in the range 0.3 -- 3~MeV on AmBe data.
\end{itemize}

The reported values were obtained by performing a Gaussian fit to the LY distribution of each population within the specified energy range. To adopt a conservative approach, the standard deviation ($\sigma$) of the distribution is quoted as uncertainty. 
For the LY of nuclear recoils, the uncertainty is larger than that of other event types, as it accounts for the combined contribution of recoils on Li, O, and W nuclei. 
Finally, defining the discrimination power (DP) as, 
\begin{equation}
    \ce{DP} = \frac{\mu_{\beta/\gamma}-\mu_{\alpha}}{\sqrt{\sigma_{\beta/\gamma}^2+\sigma_{\alpha}^2}}
\end{equation}
where $\mu$ and $\sigma$ refer to the LY distribution for the different event types, the particle identification capabilities of the \LWO\ dual-readout detector can be quantified. Using the data collected in this work, the discrimination between nuclear recoils (NR) and $\beta/\gamma$ particles is $DP = 6.8$, while between $\alpha$ and $\beta/\gamma$ particles is $DP = 7.6$. It is worth noting that these values represent an average within considered energy ranges, which vary for different event populations. This reflects the need to optimize energy ranges and to increase statistics. Further measurements will be performed to investigate particle identification as a function of energy, improving the energy threshold and the light collection efficiency to assess particle discrimination at lower energies.

\subsection{Radio- and chemical purity of the \LWO\ crystal}
\begin{figure}
    \centering
    \includegraphics[width=\linewidth]{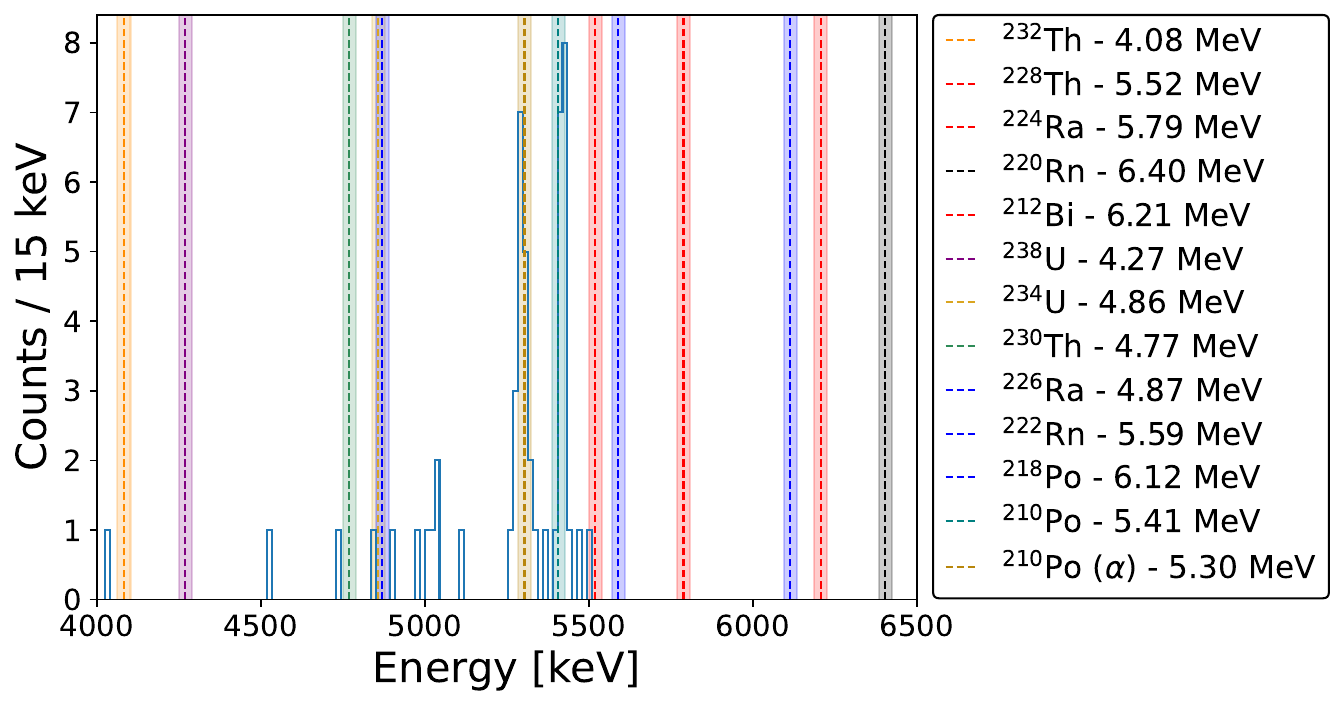}
    \caption{Energy spectrum of the $\alpha$ region acquired over 98~hours with the \LWO\ crystal. The highlighted Regions of Interest (ROIs), defined as $\pm$ 20~keV around the Q-values, are used to estimate the contamination from isotopes belonging to $^{232}$Th and $^{238}$U decay chains.}
    \label{fig:alpha}
\end{figure}

One of the most effective methods to assess the internal contamination of cryogenic calorimeters caused by radionuclides from uranium and thorium decay chains relies on analyzing the $\alpha$ region of the energy spectrum. By exploiting the extremely low counting rate in the 4--9~MeV energy range of the \LWO\ spectrum ($^{232}$Th calibration data), upper limits on the presence of natural radioactive isotopes can be set. It is assumed that the latter are homogeneously distributed within the crystal bulk, since the only evidence of surface contamination concerns $^{210}$Po due to radon exposure, for which counts are observed at $E_{\alpha} = 5300$ keV caused by the escaping nuclear recoil (107~keV).
The regions of interest (ROIs) are defined around the Q-value of each isotope as reported in Fig.~\ref{fig:alpha}.
The width of each ROI is set to $\pm$ 20~keV, corresponding to approximately 5.7$\sigma$. Figure \ref{fig:alpha} shows the energy spectrum in the $\alpha$ region, along with the highlighted ROIs corresponding to the isotopes under investigation. 
Upper limits on the contamination level are estimated under the zero-background assumption, following the Feldman-Cousins unified approach~\cite{Feldman_1998}. 
The most prominent contribution to the background spectrum originates from $^{210}$Po ($Q = 5407$ keV), which features a specific activity of 0.7 $\pm$ 0.3~mBq/kg. 
Additionally, a broader secondary peak is observed at 5304~keV, attributed to $\alpha$ decays of $^{210}$Po occurring on the crystal surface. 
In this case, the nuclear recoils can escape without releasing energy into the detector.
For all other elements, limits were set as shown in Tab.~\ref{tab:radio}.

\begin{table}[ht!]
\caption{Specific activity, in units of mBq/kg, of natural radioactive elements in the \LWO\ crystal evaluated by analyzing the $\alpha$ region of the collected data. Limits are at 90\% C.L..} 
\begin{center}
\begin{tabular}{lcc}
\hline
Chain & Nuclide  & Activity \\ 
\noalign{\smallskip}\hline\noalign{\smallskip}
$^{232}$Th & $^{232}$Th & $<$0.3\\
 & $^{228}$Th & $<$0.5\\
\noalign{\smallskip}\hline\noalign{\smallskip}
$^{238}$U & $^{238}$U & $<$0.3\\
& $^{234}$U & $<$0.3\\
& $^{230}$Th & $<$0.3\\
& $^{226}$Ra & $<$0.3\\
& $^{210}$Pb/$^{210}$Po & $0.7 \pm 0.3$\\
\noalign{\smallskip}\hline
\end{tabular}
\label{tab:radio} 

\end{center}
\end{table}

The chemical purity of the \LWO\ crystal sample was investigated using a quadrupole Inductively Coupled Plasma-Mass Spectrometer equipped with a collision cell (Agilent model 7850). A semi-quantitative analysis was performed, by calibrating the instrument with a single standard solution containing elements at a known concentration level (10~ppb of Li, Co, Y, Ce, and Tl). The contamination levels of K, Th and U were measured using a High
Resolution Inductively Coupled Plasma Mass Spectrometer (HR-ICP-MS, Thermo Fisher Scientific ELEMENT2) to enhance sensitivity. 
For K "cool plasma" mode was adopted to reduce the background due to Ar based polyatomic interferences ($^{38}$Ar$^1$H$^+$). In both cases, an uncertainty of approximately 25\% was estimated for the determined impurity concentration, while the given concentration values are expressed in parts per billion (ppb). A 21~mg \LWO\ crystal sample in a form of fine powder was placed in a plastic vial with 1~ml of ultrapure water. This mixture was then placed in an ultrasonic bath at 60$^{\circ}\mathrm{C}$ until complete sample dissolution. The resulting solution was further diluted with ultrapure water to a total volume of 10~ml and acidified with 1\% HNO$_{3}$ (a dilution factor of approximately 1500) in preparation for ICP-MS analysis. The results of the ICP-MS measurements, listed in Tab.~\ref{tab:ICPMSconcentration}, demonstrate the high chemical purity of the \LWO\ crystal sample, confirming the effectiveness of the purification process employed for the initial WO$_{3}$ powder, as well as the correctness of material handling procedures. For instance, the purification is highly effective in removing Th and U, with established upper limits below 0.4 and 0.3~ppb, respectively. 
Moreover, the \LWO\ crystal is also pure with respect to certain low-energy $\alpha$-emitting elements, such as Sm and Pt, that guarantee a low background counting rate in the nuclear recoil region. Nevertheless, the crystal contains an evident amount of Rb (135~ppb) and Cs (450~ppb), which have similar chemical properties to Li, being all related to the alkali metals, and most probably came from the Li$_{2}$CO$_{3}$ powder during the \LWO\ compound synthesis. Both these elements could be dangerous in case of rare decay searches, especially at low energies, since rubidium contains a natural $^{87}$Rb isotope (i.a. = 27.8\%, Q$_{\beta}$ = 282~keV, T$_{1/2}$ = 4.97 $\times$ 10$^{10}$ years) that decay through $\beta$-decay with a yield of 100\%. While cesium could be contaminated by an artificial $^{137}$Cs isotope, which is impossible to separate from natural cesium, and that can spoil the $\beta/\gamma$ background region up to 662~keV.
However, the most intriguing chemical contamination is molybdenum. As shown in Table~\ref{tab:ICPMSconcentration}, its concentration (8000~ppb) is the highest among all other elements. This can be explained by the similar chemical properties of Mo and W. A dedicated ICP-MS measurement was performed to determine the isotopic abundance of Mo, which is present in the \LWO\ crystal; the results are listed in Tab.~\ref{tab:IsotopeAbundance}. It is evident from the data that molybdenum does not have a natural isotopic abundance, but rather a distorted one, with the largest excess in the $^{100}$Mo isotope. This can be explained by the “growth history” of the used platinum crucible. The latter was used earlier to produce a series of enriched and natural \LMO\ crystals in the framework of R\&D activities for the CUPID experiment~\cite{CUPID:2022opf,CUPID:2025wbt,AMoRE:2024cql}. During these R\&Ds, molybdenum of a mixed isotopic abundance was embedded in the platinum crucible walls. Then, during the growth of the \LWO\ crystal, it was emanated at the melting temperature from the walls and incorporated into the crystal bulk material. Similar effect was earlier observed during production of the $^{arch}$PbMoO$_{4}$ crystal, as reported in Refs.~\cite{Pattavina:2019eox,Nagorny_2017}. This is further evidence that for the production of a low-background crystal for studying rare nuclear processes, only newly produced crucibles from refined materials should be used.

\begin{table}[ht!]
\caption{Concentration, in units of parts-per-billion (ppb), of some selected chemical impurities in the \LWO\ crystal sample as determined by ICP-MS analysis at LNGS. The uncertainty of the measured concentrations is approximately 25\%. Limits are listed at 68\% C.L..} 

\begin{center}
\begin{tabular}{lc}
\hline
Element & Concentration \\

\noalign{\smallskip}\hline\noalign{\smallskip}
K & < 2000\\
Zn & < 1000\\
Rb &  135\\
Mo &  8000\\
Cs & 450\\
Ce & 85\\
Sm & < 2\\
Pt & 10\\
Pb & 500\\
Th & < 0.4\\
U & < 0.3\\
\noalign{\smallskip}\hline
\end{tabular}
\label{tab:ICPMSconcentration} 

\end{center}
\end{table}

\begin{table}[ht!]
\caption{Isotopic abundance (in \%) of molybdenum in the \LWO\ crystal sample from the ICP-MS measurements performed at LNGS and as quoted in Ref.~\cite{CIAAW}.}
\begin{center}
\begin{tabular}{lcc}
    \hline
Isotope & Natural & As measured \\ 
        & abundance~\cite{CIAAW} & in this study\\
\noalign{\smallskip}\hline\noalign{\smallskip}
$^{92}$Mo & 14.649(106) & 14.0(7)\\
$^{94}$Mo & 9.187(33)& 8.0(4)\\
$^{95}$Mo & 15.873(30)& 14.2(7)\\
$^{96}$Mo & 16.673(8)& 15.0(8)\\
$^{97}$Mo & 9.582(15) & 8.7(4)\\
$^{98}$Mo & 24.292(80)& 22.6(1.1)\\
$^{100}$Mo & 9.744(65)& 17.5(9)\\

\noalign{\smallskip}\hline
\end{tabular}
\label{tab:IsotopeAbundance} 

\end{center}
\end{table}


\section{Conclusions and perspectives}
This work presents the first successful operation of a large-mass molybdenum-free \LWO\ crystal as a dual-readout cryogenic calorimeter. This detector, grown using the LTG Czochralski technique, which enables the production of high-quality crystals on a kilogram scale, represents a significant technological achievement in low-background searches. It is the first \LWO\ crystal to enable particle identification at milliKelvin temperatures in the keV energy range. A baseline energy resolution of \LWOBaseline keV RMS and a low-energy threshold of \LWOThreshold keV were demonstrated, which, although not yet suitable for CE$\nu$NS detection, approaches the performance observed in similar cryogenic calorimeters.

On the particle identification side, the obtained results demonstrate that \LWO\, although still not meeting the energy thresholds required for CE$\nu$NS detection, exhibits a particle discrimination capability ($>6\sigma$) at energies higher than 300~keV comparable to the best-performing detectors in rare-event physics, and better than previous \LWO\ measurements. This confirms the potential of such scintillating material as a promising candidate for low-background searches.

The excellent radiopurity of the \LWO\ crystal was also confirmed, only limits below 0.5~mBq/kg were set on the $\alpha$-active nuclides from Th/U natural decay chains, as demonstrated by the absence of visible $\alpha$ peaks in the high-energy region. The only observed $\alpha$ contamination with the activity of 0.7$\pm$0.3~mBq/kg was ascribed to $^{210}$Pb/$^{210}$Po and, most probably, caused by a surface contamination of the crystal during its machining and handling. From the other side, the crystal contains an evident amount of Rb (135~ppb) and Cs (450~ppb), which could be dangerous in case of rare decay searches in energies ranging up to 662~keV. Rubidium contains a natural $^{87}$Rb isotope (i.a. = 27.8\%, Q$_{\beta}$ = 282~keV, T$_{1/2}$ = 4.97 $\times$ 10$^{10}$ years) that decay through $\beta$ decay with a yield of 100\%. In comparison, cesium could be contaminated by an artificial $^{137}$Cs isotope, which is impossible to separate from the natural cesium isotope. Therefore, extra measures should be taken to ensure the deep purification of raw materials (especially Li$_{2}$CO$_{3}$ powder) against these two elements. As well as to avoid possible cross-contamination during the crystal growth process, \LWO\ crystals for low-background applications should be produced only in new platinum crucibles from refined materials.

The following steps include coupling this crystal to a Transition Edge Sensor (TES) and operating the setup in a low-background cryostat to characterize the low-energy spectrum and assess the presence of any excess background. 
The relatively large mass crystal (24~g) could be segmented or combined with optimized TES geometries to achieve eV-scale thresholds. Finally, the \LWO\ crystal used in this measurement could serve as an excellent monitoring tool to assess the environmental neutron background in underground cryogenic facilities.

In conclusion, the achieved performance clearly indicates that molybdenum-free \LWO\ is a favorable material for low-background and rare-event searches. Its excellent radio-purity, dual-readout capabilities, and neutron sensitivity via $^6$Li(n,$\alpha$)$^3$H make it a versatile candidate for several interesting applications. A detailed background measurement campaign in a source-free configuration is planned to explore its potential further.

\section*{Acknowledgements} 

We are particularly grateful to Dr. J. Walker Beeman for his invaluable support in the production of NTDs. We are also grateful to M. Guetti for the assistance in the cryogenic operations and the mechanical workshop of LNGS. This work was supported by the Italian Ministry of University and Research (MUR) through the grant Progetti di ricerca di Rilevante Interesse Nazionale (PRIN) grant no. 2022L2AXP2. This material is also based upon work supported by the European Union’s Horizon 2020 research and innovation program under the Marie Skłodowska-Curie Grant Agreement No. 101029688 (ACCESS—Array of Cryogenic Calorimeters to Evaluate Spectral Shape). This work was supported by the Ministry of Science and Higher Education of the Republic of Kazakhstan (Grant No. AP19676308) and Nazarbayev University (Grant No. 20122022FD411). The research of V.N.S. and V.D.G. was supported by the Ministry of Science and Higher Education of the Russian Federation, grant N. N121031700314-5.

\section*{Availability of data and material}
Data will be made available on reasonable request to the corresponding author.

\bibliography{main.bib}
\bibliographystyle{spphys} 

\end{document}